\documentclass[twocolumn,showpacs,prb,superscriptaddress]{revtex4}

\usepackage{graphicx}

\newcommand{\av}{_{\mathrm{av}}}
\newcommand{\nsw}{N_{\mathrm{sweep}}}
\newcommand{\nsa}{N_{\mathrm{samp}}}
\newcommand{\figurewidth}{\columnwidth}

\begin{document}

\title{Correlation length of the two-dimensional Ising spin glass with 
bimodal interactions}

\author{Helmut G.~Katzgraber}
\affiliation
{Theoretische Physik, ETH H\"onggerberg,
CH-8093 Z\"urich, Switzerland}

\author{L.~W.~Lee}
\affiliation{Department of Physics,
University of California,
Santa Cruz, California 95064}

\date{\today}

\begin{abstract}
We study the correlation length of the two-dimensional Edwards-Anderson Ising
spin glass with bimodal interactions using a combination of parallel tempering
Monte Carlo updates and a rejection-free cluster algorithm in order to 
speed up equilibration. Our results show that the correlation
length grows $\sim \exp(2J/T)$ suggesting through hyperscaling
that the degenerate ground state is separated from the first excited
state by an energy gap $\sim 4J$, as would naively be expected.
\end{abstract}

\pacs{75.50.Lk, 75.40.Mg, 05.50.+q}
\maketitle

\section{Introduction}
\label{sec:introduction}

Despite intense research,\cite{binder:86,mezard:87} spin glasses still pose
many unanswered questions. Thus it is of great interest to understand the 
nature of the spin-glass state for realistic short-range spin glasses. A model
that has proven to be a good ``workhorse'' due to its simplicity and ease
of implementation is the Edwards-Anderson Ising spin glass\cite{edwards:75} 
with bimodal interactions.
Here we consider this model in two dimensions, for which the spin-glass
transition occurs at zero temperature. Nonetheless the
low-temperature properties of the model remain controversial: There are
different predictions on the energy gap between the ground state and the 
first excited state, and the critical behavior of the correlation length is
not fully understood.

Wang and Swendsen\cite{wang:88} first suggested an exponential scaling of the
specific heat of the two-dimensional bimodal Ising spin glass:
\begin{equation}
C_{V} \sim T^{-2} e^{-A \beta J} \; ,
\end{equation}
where $\beta = 1/T$ represents the inverse temperature, $J$ is the mean
interaction strength of the (random) bonds, and $A$ is a numerical prefactor.
Although they argue analytically that the energy gap should be $4J$, i.e., $A
= 4$, their final estimate from their numerical work is $A = 2$. In addition,
they estimated the correlation length exponent $\nu$ to be finite and
positive, implying power-law scaling of the correlation length. Unfortunately,
their simulations were done on small system sizes with few disorder
realizations, suggesting that their results are governed by corrections to 
scaling.

Saul and Kardar,\cite{saul:93} by means of an exact integer algorithm
to estimate the partition function of the two-dimensional $\pm J$ spin
glass, later argued that $A = 4$ and, in addition, proposed an exponential
scaling form for the correlation length, i.e.,
$\xi \sim \exp(n \beta J)$ with $n =2$. The
exponential scaling present in the bimodal glass was later verified by
Houdayer\cite{houdayer:01} by analyzing the scaling properties of the
spin-glass susceptibility and the Binder cumulant\cite{binder:81} for large
system sizes using a novel cluster algorithm. Recently, Lukic {\em et
al.}\cite{lukic:04}~reconsidered the problem by computing the exact partition
function of the system for system sizes up to $50 \times 50$ spins. They
conclude that $A = 2$. According to hyperscaling, the singular part of the 
free energy scales as $\xi^{-2}$ (in two dimensions) and so, if $\xi$ diverges
exponentially, the leading exponential dependence of the specific heat will 
also be given by  $\xi^{-2}$. Therefore, one expects $A = 2 n$, and in the
case of Lukic {\em et al.}~this would mean that $\xi \sim \exp(n \beta J)$
with $n = 1$. For a summary of all predictions of $A$ and $n$, the 
prefactors to $\beta J$ in the exponential scaling of specific heat and 
correlation length, respectively, see Table \ref{exponents}.

\begin{table}
\caption{
Different estimates for $n$ and $A$.
From finite-size scaling arguments in two dimensions, we expect that the
specific heat scales as $C_{V} \sim \xi^{-2}$ where $\xi$ is the correlation
length. Following the exponential scaling for $C_{V}$ suggested by Saul and
Kardar (Ref.~\onlinecite{saul:93}), we expect in general 
$C_{V} \sim \exp(-A \beta J)$ and $\xi \sim \exp(n \beta J)$, with $A = 2n$.
\label{exponents}
}
\begin{tabular*}{\columnwidth}{@{\extracolsep{\fill}} l l l }
\hline
\hline
Reference &  $n$  & $A$ \\
\hline
Wang and Swendsen (Ref.~\onlinecite{wang:88})   &    & 2  \\
Saul and Kardar (Ref.~\onlinecite{saul:93})	& 2  & 4  \\
Houdayer (Ref.~\onlinecite{houdayer:01})        & 2  &    \\
Lukic {\em et al}.~(Ref.~\onlinecite{lukic:04}) & 1  & 2  \\
Katzgraber and Lee (this work)			& 2  &    \\
\hline
\hline
\end{tabular*}
\end{table}

In this work we compute the finite-size correlation 
length\cite{cooper:82,ballesteros:00} directly via Monte Carlo simulations for large 
system sizes and show that 
$n$ changes continuously for intermediate system sizes, yet saturates at $n
\approx 2$ for large enough systems. This suggests via hyperscaling 
that $A = 4$, i.e., the energy gap in the two-dimensional Ising spin glass is
$\sim 4 J$.

The paper is structured as follows. In Sec.~\ref{sec:model} we introduce the
model and observables and in Sec.~\ref{sec:results} we present our 
results. Conclusions are summarized in Sec.~\ref{sec:conclusions}.

\section{Model and Observables}
\label{sec:model}

The Hamiltonian of the two-dimensional Ising spin glass is given by
\begin{equation}
{\cal H} = - \sum_{\langle i,j\rangle} J_{ij} S_i S_j ,
\label{eq:ham}
\end{equation}
where the sum ranges over nearest neighbors on a square lattice with periodic
boundary conditions. $S_i$ represent Ising spins taking values $\pm 1$, and
the interactions $J_{ij}$ are bimodally distributed, i.e., $J_{ij} \in {\pm 1}$.
For the Monte Carlo simulations we use a combination of single-spin flips,
parallel tempering updates,\cite{hukushima:96,marinari:98b} and rejection-free 
cluster moves\cite{houdayer:01,cluster_algorithm} in order to speed up 
equilibration. 
To ensure that the system is equilibrated, we perform a logarithmic data
binning of all observables (energy, spin-glass susceptibility, and correlation
length) and require that the last three bins agree within
error bars and are independent of the number of Monte Carlo sweeps $\nsw$.
The parameters of the simulation are listed in Table 
\ref{simparams}. 
\begin{table}
\caption{
Parameters of the simulations. $\nsa$ represents the number of disorder
realizations computed; $\nsw$ the total number of Monte Carlo sweeps
of the $2 N_T$ replicas for a single sample. $N_T$ is the number of
temperatures in the parallel tempering method and $T_{\rm min}$ represents the
lowest temperature simulated. For $L \le 16$ we have used standard parallel
tempering Monte Carlo simulations, whereas for $L \ge 24$ we have used a 
combination of parallel tempering and cluster updates 
(Ref.~\onlinecite{houdayer:01}).
\label{simparams}
}
\begin{tabular*}{\columnwidth}{@{\extracolsep{\fill}} c r r r l }
\hline
\hline
$L$  &  $\nsa$  & $\nsw$ & $T_{\rm min}$ & $N_{\rm T}$  \\
\hline
  4 & $10\; 000 $ & $2.0 \times 10^5$ & 0.050 & 20 \\
  6 & $10\; 000 $ & $2.0 \times 10^5$ & 0.050 & 20 \\
  8 & $10\; 000 $ & $2.0 \times 10^5$ & 0.050 & 20 \\
 12 & $10\; 000 $ & $4.0 \times 10^5$ & 0.050 & 20 \\
 16 & $10\; 000 $ & $1.0 \times 10^6$ & 0.050 & 20 \\
 24 & $ 5\; 003 $ & $2.0 \times 10^6$ & 0.050 & 20 \\
 32 & $ 5\; 031 $ & $2.0 \times 10^6$ & 0.050 & 20 \\
 64 & $     500 $ & $4.2 \times 10^6$ & 0.200 & 39 \\
 96 & $     609 $ & $6.5 \times 10^6$ & 0.200 & 63 \\
128 & $     420 $ & $2.0 \times 10^6$ & 0.396 & 50 \\
\hline
\hline
\end{tabular*}
\end{table}

As mentioned in Sec.~\ref{sec:introduction}, according to Saul and 
Kardar,\cite{saul:93} the correlation length of the 
two-dimensional $\pm J$ spin glass scales as
\begin{equation}
\xi \sim e^{n\beta J} \; ,
\label{eq:expscale}
\end{equation}
where $\beta = 1/T$ and $n=2$. In this paper we determine the value of $n$
from Monte Carlo simulations.

In order to compute the correlation length $\xi$, as well as
estimate the critical exponent $\eta$, we compute the spin-glass susceptibility
\begin{equation}
\chi_{\mathrm{SG}} = N [\langle q^2 \rangle]_{\av} \; ,
\label{eq:chi}
\end{equation}
where $[\cdots]_{\av}$ represents a disorder average and $\langle \cdots
\rangle$ a thermal average, and
\begin{equation}
q = \frac{1}{N} \sum_{i = 1}^{N}S_i^\alpha S_i^\beta \; 
\label{eq:q}
\end{equation}
is the Edwards-Anderson spin-glass order parameter.
In Eq.~(\ref{eq:q}) $N = L^2$ represents the number of spins and $\alpha$
and $\beta$ represent two replicas of the system with the same disorder. 
Due to the exponential scaling\cite{saul:93} of the correlation length we
expect\cite{houdayer:01}
\begin{equation}
\chi_{\mathrm{SG}}(T,L) = L^{2-\eta}\tilde{C}\left[\beta - 
\frac{1}{n}\ln L\right] \;.
\label{chiscale1}
\end{equation}
The finite-size correlation
length\cite{cooper:82,kim:94,palassini:99b,ballesteros:00,lee:03,katzgraber:04} 
$\xi_L$ is given by
\begin{equation}
\xi_L = {1 \over 2 \sin (|{\bf k}_\mathrm{min}|/2)}
\left[{\chi_{\mathrm{SG}}(0) \over
\chi_{\mathrm{SG}}({\bf k}_\mathrm{min})} - 1 \right]^{1/2} \; ,
\label{eq:xiL}
\end{equation}
where ${\bf k}_\mathrm{min} = (2\pi/L, 0, 0)$ is the smallest nonzero
wave vector, and $\chi_{\mathrm{SG}}({\bf k})$ is the  wave-vector-dependent
spin-glass susceptibility:
\begin{equation}
\chi_{\mathrm{SG}}({\bf k}) = {1 \over N} \sum_{i,j} [\langle S_i S_j
\rangle^2 ]\av e^{i {\bf k}\cdot({\bf R}_i - {\bf R}_j) } \; .
\label{eq:chik}
\end{equation}
Because $\xi_L/L$ is dimensionless, we
expect\cite{palassini:99b,ballesteros:00,lee:03,katzgraber:04}
\begin{equation}
{\xi_L / L } = \tilde{X} \left[ \beta - \frac{1}{n} \ln L \right]
\;\;\;\;\;\;\;\;\;\;\;(T_{\rm c} = 0) \; .
\label{eq:fss}
\end{equation}
Here $\tilde{X}$ is a scaling function.
Note that in Eqs.~(\ref{chiscale1}) and (\ref{eq:fss}) we have explicitly 
left $n$ as a ``variable.'' 

To determine the asymptotic value of $n$, we compute the {\em bulk} correlation 
length $\xi_\infty$ at low temperatures using the method of 
Kim\cite{kim:94} (first introduced in Ref.~\onlinecite{luscher:91}). 
The finite-size scaling
relation for the correlation length can be written as
\begin{equation}
{\xi_\infty \over L} = f\left({\xi_L \over L}\right) \, ,
\label{gx}
\end{equation}
where we determine $f(x)$ by fitting to data in the range
$0.500 < T < 1.391$ where we have data for the correlation length in
both the bulk and finite-size regimes. Using $f(x)$, we then determine
$\xi_\infty$ from Eq.~(\ref{gx}) using
data for $L=96$ and $128$ in the range $ 0.396  \le T \le 1.391$.

\section{Results}
\label{sec:results}

\begin{figure}
\includegraphics[width=\figurewidth]{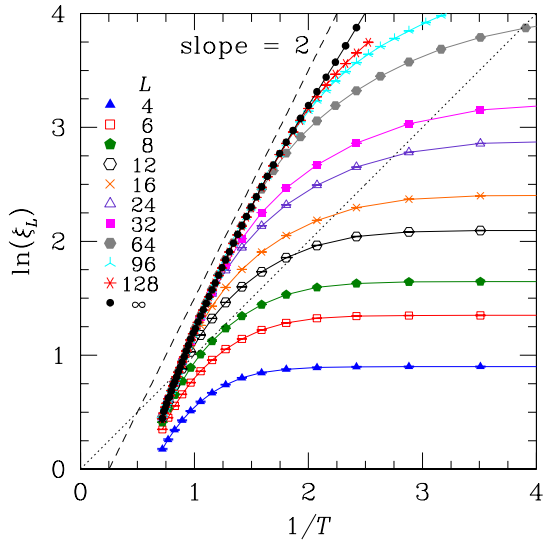}
\vspace*{-1.0cm}
\caption{
(Color online)
Data for the finite-size correlation length: $\ln(\xi_L)$ vs $1/T$. The slope
of the curves in the thermodynamic limit determine $n$.
For small values of $1/T$ the data are independent of system size $L$, but 
the data ``peel off'' from this common curve at a value of $1/T$ which
increases with increasing size. The slope of the common curve 
asymptotically seems to approach the value $n = 2$ (dashed line). 
This is supported by data for the bulk correlation length
($\bullet$) which
shows an asymptotic slope also compatible with $\xi \sim \exp(2\beta J)$,
i.e., $n = 2$. (The dotted diagonal line has slope 1.)
}
\label{fig:lxil}
\end{figure}

In Fig.~\ref{fig:lxil} we show data for the natural logarithm of 
the finite-size correlation length 
as a function of $1/T$ for different system sizes. According to
Eq.~(\ref{eq:expscale}), we expect data for $\ln(\xi_L)$ vs $1/T$ to
asymptotically approach a slope of $n$. The data presented in
Fig.~\ref{fig:lxil} show good agreement with $n \approx 2$, which in turn
agrees with the results of Refs.~\onlinecite{saul:93} and
\onlinecite{houdayer:01}. 
We have also used the extrapolation
method by Kim\cite{kim:96} and show data for the bulk correlation length
extrapolated from data for the largest system sizes. 
These data also agree well with 
the naive scaling with $n = 2$, and not with $n = 1$ (dotted line in
Fig.~\ref{fig:lxil}), as predicted by
Lukic {\em et al.}\cite{lukic:04} from specific heat studies.
Corrections to scaling in the two-dimensional bimodal
spin glass are strong for system sizes $L$ up to $\sim 64$. Hence the results
of Lukic {\em et al.} for $L \leq 50$ could be influenced by corrections to scaling 
that mask the true critical behavior. 

\begin{figure}
\includegraphics[width=\figurewidth]{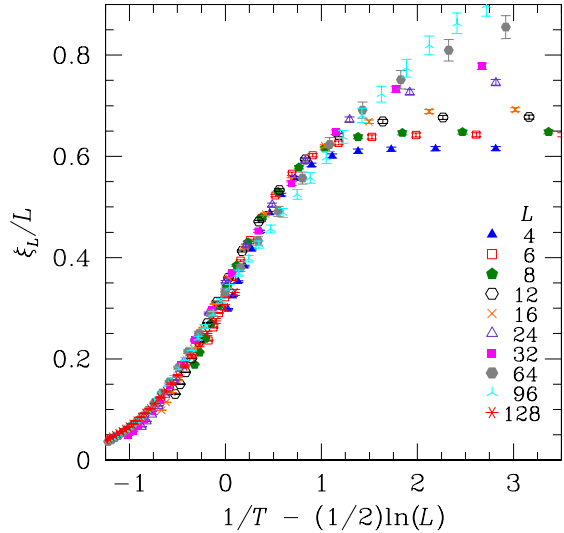}
\vspace*{-1.0cm}
\caption{
(Color online)
Parameter-free scaling plot of the finite-size correlation length according to
Eq.~(\ref{eq:fss}) using $n = 2$. Only the data for large values of $L$ 
($L \gtrsim 64$) scale for low temperatures. 
}
\label{fig:scalexi}
\end{figure}

\begin{figure}
\includegraphics[width=\figurewidth]{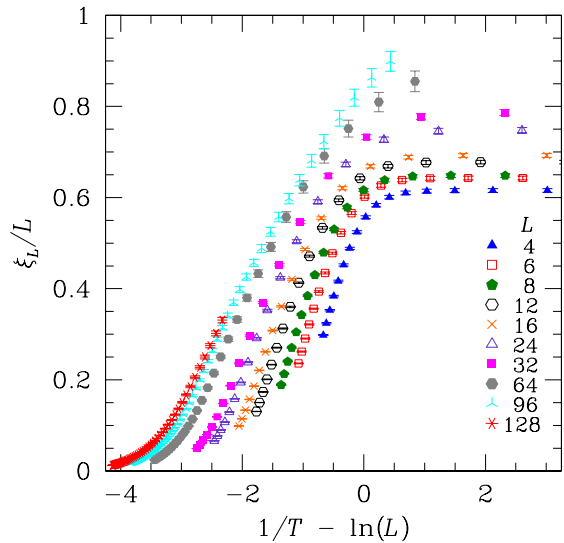}
\vspace*{-1.0cm}
\caption{
(Color online)
Parameter-free scaling plot of the finite-size correlation length according to
Eq.~(\ref{eq:fss}) using $n = 1$. The data do not scale well for any
temperature and system size.
}
\label{fig:scalexinone}
\end{figure}

In order to test this hypothesis, we show a finite-size scaling plot of the
data of the finite-size correlation length according to 
Eq.~(\ref{eq:fss}) with $n = 2$ (Fig.~\ref{fig:scalexi}). We see that 
the data for $L \lesssim 64$ are not in the asymptotic limit, whereas the 
data for $L \gtrsim 64$ scale reasonably well. This shows that in order 
to obtain precise estimates of $n$, very large system sizes are 
required.\cite{campbell:04}
In addition, the exponential scaling means that, for the two-dimensional
bimodal spin glass, the critical exponent for the correlation length is
infinite, i.e., 
\begin{equation}
\nu = \infty \; .
\label{eq:nu}
\end{equation}
We have also tried to scale the data according to Eq.~(\ref{eq:fss}) using $n
= 1$ in accordance with the prediction of Lukic {\em et al}.\cite{lukic:04}
As shown in Fig.~\ref{fig:scalexinone} the data do not scale well for any
system size and any temperature suggesting that $n = 1$ is not probable.

\begin{figure}
\includegraphics[width=\figurewidth]{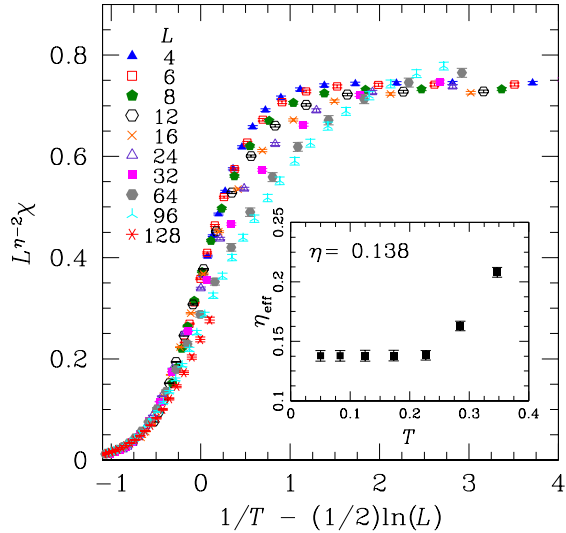}
\vspace*{-1.0cm}
\caption{
(Color online)
Scaling plot of the spin-glass susceptibility with $\eta = 0.138$ and $n = 2$. 
The inset shows the effective critical exponent $\eta_{\rm eff}(T)$.
}
\label{fig:chisg}
\end{figure}

Data for the spin-glass susceptibility show similar finite-size effects as in
the case of the correlation length. In Fig.~\ref{fig:chisg} we show a
finite-size scaling plot of the spin-glass susceptibility $\chi_{\rm SG}$
assuming $n = 2$ and $\eta = 0.138$. 

Note that scaling works ``reasonably
well'' for a large range of $\eta$ values (not shown) thus illustrating
again\cite{katzgraber:02a}
that the spin-glass susceptibility is not a good observable with which
to study critical
properties of glassy systems.

At $T = T_{\rm c}$ $(= 0)$ we expect, according to Eq.~(\ref{chiscale1}), 
\begin{equation}
\chi_{\rm SG} \sim L^{2 - \eta} \;\;\;\;\;\;\;\;\; (T = T_{\rm c})\; .
\label{eq:chitc}
\end{equation}
In order to estimate the critical exponent $\eta$, we plot 
$\ln(L^{-2}\chi_{\rm SG})$ vs $\ln(L)$ and extract the slope of the curves,
which should be proportional to a temperature-dependent effective exponent
$-\eta_{\rm eff}(T)$, where
\begin{equation}
\lim_{T \rightarrow T_{\rm c} = 0} \eta_{\rm eff}(T) = \eta \; .
\label{eq:etalimes}
\end{equation}
The inset of Fig.~\ref{fig:chisg} shows the effective exponent 
$\eta_{\rm eff}(T)$ as a function of temperature. We estimate from our data
\begin{equation}
\eta = 0.138 \pm 0.005 \;,
\end{equation}
a result which is slightly smaller than a recent estimate by 
Houdayer.\cite{houdayer:01} 
We expect the value of the critical exponent $\eta$
to change slightly with larger system sizes.
However, because we do not have data for the largest $L$ 
at the lowest temperatures
(see Table \ref{simparams}), we cannot make a strong statement 
regarding $\eta$.

\section{Conclusions}
\label{sec:conclusions}

To conclude, we have shown that the correlation length of the two-dimensional
Ising spin glass with bimodal interactions scales exponentially 
$\sim \exp(n \beta J)$ with $n = 2$. These results are in
agreement with previous work by Saul and Kardar\cite{saul:93}
and Houdayer.\cite{houdayer:01}
However, if one assumes the hyperscaling result $A = 2 n$, our results
disagree with specific heat studies by Lukic {\em et al.}\cite{lukic:04}
as well as Wang and Swendsen\cite{wang:88} who find $A = 2$. 
Using the hyperscaling relation, our results also imply that the
excitation gap for the bimodal spin glass is $\approx 4J$.
Exponential scaling of the correlation length means that the critical
exponent for the correlation length is infinite ($\nu = \infty$)
for the two-dimensional bimodal spin glass. In addition, we estimate the
critical exponent $\eta$ and find that $\eta \approx 0.138(5)$.

Although in this work we were able to present data that can be extrapolated
to the bulk regime for a certain temperature range therefore allowing us to
draw conclusions in the thermodynamic limit for the scaling of the correlation
length, finite-size effects are very strong in this system and so an 
analysis with yet larger system sizes at lower temperatures is desirable. 
While it is unlikely, a change in the asymptotic behavior at larger $L$ 
cannot be ruled out completely.

\begin{acknowledgments}

We would like to thank I.~A.~Campbell for helpful discussions and
A.~P.~Young for discussions and comments, as well as
critically reading the manuscript. L.W.L.~acknowledges support
from the National Science Foundation under NSF Grant No.~DMR 0337049.
The simulations were performed in part on the Asgard cluster
at ETH Z\"urich. We would also like to thank G.~Sigut and M.~Troyer for
allowing exclusive access during two weeks to the Hreidar cluster 
at ETH Z\"urich, without which these calculations would not have been 
possible.

\end{acknowledgments}

\bibliography{refs,comment}

\end{document}